# CloudLaunch: Discover and Deploy Cloud Applications

*Enis Afgan[a], Andrew Lonie[b], James Taylor[a], Nuwan Goonasekera[b]\**

[a] *Department of Biology, Johns Hopkins University, 3400 N Charles Street, Baltimore, MD 21210, USA*
[b] *Melbourne Bioinformatics, University of Melbourne, 700 Swanston Street, Carlton, VICTORIA 3053, Australia*

ABSTRACT

Cloud computing is a common platform for delivering software to end users. However, the process of making complex-to-deploy applications available across different cloud providers requires isolated and uncoordinated application-specific solutions, often locking-in developers to a particular cloud provider. Here, we present the CloudLaunch application as a uniform platform for discovering and deploying applications for different cloud providers. CloudLaunch allows arbitrary applications to be added to a catalog with each application having its own customizable user interface and control over the launch process, while preserving cloud-agnosticism so that authors can easily make their applications available on multiple clouds with minimal effort. It then provides a uniform interface for launching available applications by end users across different cloud providers. Architecture details are presented along with examples of different deployable applications that highlight architectural features.

## 1. Introduction

As cloud technologies and platforms become more mature, modern Infrastructure-as-a-Service (IaaS) clouds are broadly converging in terms of functionality and scope [1]. Nevertheless, as others have noted elsewhere [2], subtle differences in providers can easily lead to vendor lock-in. This is a significant problem in both academic and commercial settings, where heterogeneity in resource access, funding models, and geography can make it difficult to share cloud applications developed in one setting, with cloud infrastructure running in a different setting. As clouds increasingly become the de facto means of software delivery to end-users, it becomes just as important to support multiple cloud infrastructures operated by different vendors and communities.

There has been a proliferation of private and community clouds; in the academic community, there are a number of national-scale academic community clouds including, the NeCTAR cloud in Australia, the ELIXIR cloud in Europe, the Jetstream cloud [3] in the US, the CLIMB cloud [4] in the UK, and efforts in Canada, South Africa, and others. Many vendors have emerged in the commercial space with some stable big providers such as Amazon, Google, and Microsoft.

Users tend to be generally tied to whatever cloud infrastructure their hosting company or funding model allows them access, or on whichever cloud their data happens to reside. Despite the potential promise of ubiquitous cloud access, the reality is that users are often siloed within their institutional cloud, with no way to marry software available on a particular cloud, with the data and resources they have access to in their institutional cloud.

As a result, the burden increasingly falls upon cloud application developers to make their applications available on multiple clouds. This is somewhat analogous to how desktop application developers must support different operating systems to reach their target audiences. This requires that the developer:

1. Build and test their application against multiple clouds;
2. Provide a means by which a user can discover and launch an application on a cloud infrastructure of their choice;
3. Ensure that the application is orchestrated, monitored, and managed on the target infrastructure.

In previous work on CloudBridge [1], we addressed the first issue, noting that many existing solutions to the problem, such as Apache Libcloud [5] and Ansible [6], do not provide a unified abstraction for IaaS clouds, requiring that developers test their applications against individual cloud infrastructures. It is only at a level higher that containerization frameworks such as Kubernetes and Docker Swarm have alleviated this problem somewhat by letting developers deal with a Platform-as-a-Service (PaaS) level of abstraction.

In this paper, we discuss our work in addressing the second aspect of this problem, building upon previous work done in BioCloudCentral [7]. For this purpose, we built CloudLaunch - a web portal and an API platform for discovering and launching applications on multiple cloud infrastructures. Novelties introduced by CloudLaunch include the ability to describe an application once using open technologies and have it uniformly deployable on multiple cloud infrastructure providers using a web interface or a REST API. With the API driven approach, CloudLaunch can be used as a deployment engine for external applications to provide cloud abstraction and orchestration capabilities. In this context, we identify at least four potential beneficiaries of the CloudLaunch science gateway:

1. End users who want to easily discover and launch applications on multiple clouds;
2. Application deployers who want to make applications available on multiple clouds through CloudLaunch's centralized catalog;
3. Application developers who need an API-driven deployment engine to use within a custom application, say for constructing a higher-level science gateway;
4. Institutions that want to have their own catalog of applications for internal users.

## 2. An overview of CloudLaunch

For an end-user, the initial entry-point to CloudLaunch is a web interface for browsing a catalog of appliances (Fig. 1A). An appliance represents a deployable system, which can be as simple as an operating system running on a virtual machine or as complex as a virtual laboratory (e.g., GVL [8]). A key benefit of an appliance is that it comes with 'batteries included' - it provides the necessary infrastructure, applications, and configurations to deliver a functional system to the user. Next, CloudLaunch is a deployment platform that allows users to instantiate a chosen appliance on any one of a range of cloud providers via the web interface or its REST API (Fig. 1B).

*\* Corresponding author*

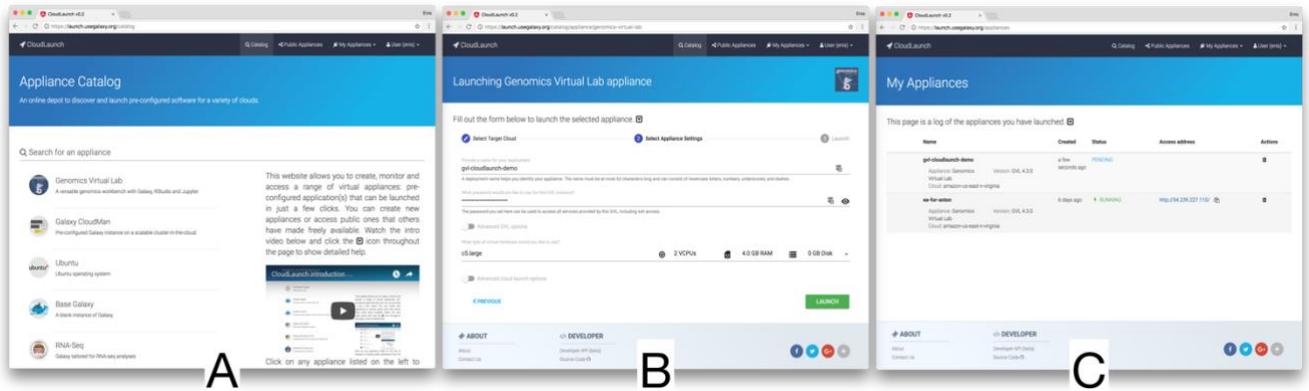

**Fig. 1.** Functionality available through CloudLaunch: (A) browse a catalog of appliances; (B) launch a chosen appliance, selecting from a range of cloud providers and launch properties; and (C) a dashboard of launched appliances showing the current status of the application deployment.

The deployment process instantiates the necessary resources on the given cloud provider and creates an environment necessary to start the software making up the appliance. Each appliance can implement its own launch logic and hence create the appropriate environment (see Section 4). Finally, CloudLaunch is a dashboard for managing launched appliances, such as checking the health status of an appliance or deleting it (Fig. 1C). Overall, the CloudLaunch application acts as a gateway for launching applications into a cloud.

The key differentiators of CloudLaunch are its uniform interface toward multiple infrastructure providers and support for integration of arbitrary applications. While other application repository services exist from a number of vendors and academic institutions (e.g., AWS Marketplace [9], Google Launcher [10], CyVerse Atmosphere [11]), none have uniform support for multiple cloud providers. They each offer a different and proprietary set of application deployment recipes to be launched on their respective infrastructure. This is undesirable for the end users because they must manually search for the desired application across multiple repositories, switch between the providers, and deal with non-uniform interfaces. For the application developer or deployer that wishes to make their application available for deployment, they must learn a proprietary format for adding their application to a repository - assuming such functionality is even supported (e.g., Google Compute Engine does not support public custom images) - and do so multiple times.

Instead, CloudLaunch appliances are based on open, well-documented, and cross-cloud technologies. This allows anyone to integrate their cloud application into a deployable appliance and do so once for multiple infrastructure providers. Because of the cloud abstraction layer implemented in CloudLaunch (via the CloudBridge library [1]), multi cloud functionality can be confidently used across multiple providers. This implies that an appliance deployment does not need to be explicitly tested against each provider yet it will dependably work. Simultaneously, end users experience a consistent process for launching any appliance on any supported infrastructure. While it may appear that users get locked into CloudLaunch instead, we have done our best to preempt such a situation by keeping its plug-in interface extremely simple, so that appliances could be easily ported to different technologies, and existing technologies can be easily wrapped into a CloudLaunch appliance. In future, we also plan to support other existing technologies such as the TOSCA standard [12].

## 3. Architecture

The design of CloudLaunch is focused on flexibility and extensibility. By flexibility, we mean the ability to support a range of usage scenarios. This can mean usage via the web frontend or the API as well as the ability to integrate with a variety of infrastructure providers or applications. While we focus on deploying applications to IaaS cloud infrastructures here, the application is designed such that appliances can be launched on other infrastructures as well, for example cloud container services or HPC clusters (under the assumption that the appropriate deployment implementation is provided). By extensibility, we mean the ability to support arbitrary appliances, at the user interface level as well as the launch process level. For example, a basic virtual machine (VM) with just an operating system requires only the appropriate firewall ports to be opened (e.g., *ssh*) and an instance to be launched. An appliance with a web interface, an FTP server, additional system users, persistent storage, etc. would require additional user interface elements (such as selecting the size of persistent storage) and a more complex launch-process actions to take place (e.g., creating a block store volume, attaching it to an instance, and formatting it as a file system).

As previously discussed, CloudLaunch currently focuses on deploying appliances on IaaS clouds. The appliance deployment process is captured in a CloudLaunch plugin as native Python code with an option to make external calls to deployment tools such as Ansible. The deployment process can be as complex as required by the appliance; for example, it can capture the simple creation of a virtual machine with just an operating system or create a complex runtime environment with a cluster manager and attached storage. The complexity of a deployment is entirely captured within the appliance plugin while CloudLaunch orchestrates steps required to run the plugin. In addition to the launch process, CloudLaunch plugins implement three additional actions: health check, restart, and delete. The health check task can perform basic checks of liveliness of a virtual machine or a complex query of the deployed application to make sure necessary services are operating as required. The restart task can implement a controlled and/or partial reboot of the system, including restarting any containers or the host virtual machine. Similarly, the delete task performs the appropriate termination of the launched appliance.

CloudLaunch does not attempt to operate outside of these boundaries. For example, CloudLaunch does not manage the deployed applications to offer runtime environment reconfiguration, perform scaling steps, or handle ongoing storage management. Those features are part of the third issue identified in the "Introduction" section and left as future work. Further, of particularly relevance to multi-cloud deployments, it is important to note that CloudLaunch focuses only on the deployment process; it requires that the required appliance resources (e.g., machine image, file system) be available on target clouds. This may require the application deployer to build necessary resources on every cloud or rely on containerization technologies to make the appliance more portable.

Technically, CloudLaunch is implemented as separate front- and back-end applications. The front-end layer interfaces with the back-end through a REST API. The back-end is structured around a core framework that brokers the interactions between requests and appliances. The persistent data (e.g., clouds, appliances) are stored in the model while the appliance logic runs as asynchronous, distributed tasks using a Celery task queue. Appliances being deployed via CloudLaunch are treated as standalone plugins. This high-level architecture is captured in Fig. 2. We have implemented CloudLaunch in Python using the Django web framework [13] with the Django REST Framework plugin [14], so that the REST API is a browsable, self-documenting interface. The front-end is implemented as a Single Page Application using the Angular framework and is written in TypeScript. At deployment, the front-end is entirely static content and can be effectively served using a scalable web server, such as Nginx. Long running, asynchronous back-end tasks are farmed to a task queue, implemented using Celery [15]. As already mentioned, interfacing with multiple cloud providers is accomplished through the CloudBridge library.

The modularity of the application hinges on the separation between the front-end, the back-end, and the appliance plugins. This leads to a simple yet flexible and powerful model for encapsulating appliances while decoupling the front- and back-ends. It further allows other applications to build custom interfaces and communicate with CloudLaunch via the backend API directly. Details of each are discussed next.

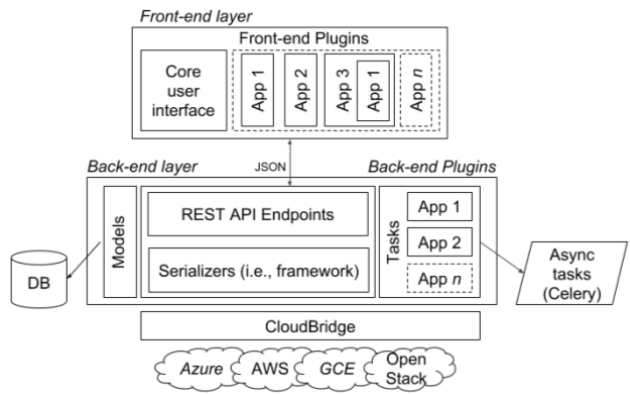

**Fig. 2.** High-level architecture of CloudLaunch. Along with the framework components that provide the core functionality, CloudLaunch is extensible via composable application plugins on the user interface and back-end. Note that the composition capability of back-end plugins is not captured in this figure.

### 3.1. Front-end Layer

The frontend is in charge of obtaining necessary information from the user while the back-end processes the information and interfaces with the infrastructure provider via the appliance plugins (Fig. 3). Each appliance is implemented as a self-contained plugin. The front-end communicates with the back-end over a REST API, passing application configuration as a flexible dictionary data structure of arbitrary complexity. Each appliance front-end needs to produce this appliance-specific JSON data structure containing any information deemed necessary for the back-end to perform its functions, which is duly conveyed by CloudLaunch to the appliance-specific back-end plugin component for processing.

#### 3.1.1. Front-end Plugin
On the front-end, an appliance is implemented as an Angular component and the UI elements are hence reusable outside CloudLaunch. To integrate with the CloudLaunch framework, the plugin component needs to implement the interface defined in Fig. 4. A `BasePluginComponent` exists in the framework that can be used as a base class for providing much of the core functionality. In return, a most basic plugin implementation can extend that base class and provide an implementation with just one method, `configName`, to indicate the top-most element of the appliance JSON data structure. Appliances with custom launch forms also need to implement a form group that captures the necessary data from the user interface.

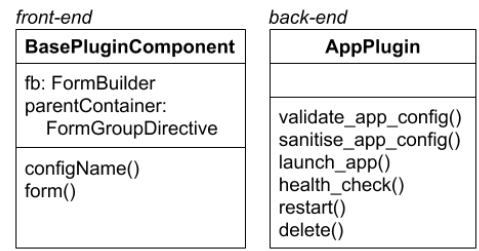

**Fig. 2.** Plugin interfaces that need to be implemented by each appliance plugin for the front-end and back-end.

### 3.2. Back-end Layer

The back-end of the framework is exposed through a self-documenting, browsable REST API. The API defines four top level endpoints: *applications*, *infrastructure*, *deployments*, and *authentication*. Resources exposed via the infrastructure endpoint map to cloud provider IaaS resources uniformly through CloudBridge, allowing for the manipulation of the underlying IaaS resources. Therefore, the structure of the infrastructure endpoint closely mirrors that of CloudBridge and provides nested endpoints for browsing or creating compute, storage, security, and network resources. Note that this endpoint is completely decoupled from CloudLaunch itself, and can be used irrespective of the rest of the CloudLaunch functionality; it is available as a separate, pip-installable API for interfacing with cloud providers, maximizing reusability and modularity (https://github.com/CloudVE/djcloudbridge).

The remaining endpoints (applications, deployments, and authentication) interact with the CloudLaunch database to expose and manage stored information. The applications endpoint represents the catalog of registered appliances, each including the necessary linkage to realize the concept of an appliance (e.g., clouds where the application is available with relevant machine images, required security group and ports, a pointer to which backend plugin to use for the appliance as well as a pointer to which front-end plugin to use, launch properties, etc.). The deployments endpoint lists appliance deployments that were performed by the user. It captures details such as which cloud the appliance was launched on, what launch properties were used and what output the deployed appliance produced, such as the instance IP address or the URL through which to access the deployed appliance. On each deployment, it is possible to initiate tasks. These tasks perform health check actions, restart, or delete actions on that deployment. The list of tasks can also be extended as the need arises in the future. Finally, the authentication endpoint handles user registration and credential safekeeping. The user registration is handled via Django Social Auth library, which allows for pluggable authentication mechanisms. The cloud credentials are stored in the database as encrypted fields. The encryption is handled directly in the database using fernet keys, which allow for regular key rotation.

### 3.2.1. Back-end Plugin

The back-end appliance plugin needs to provide an implementation for the interface defined in Fig. 4. The core functionality provided by the plugin is parsing the JSON data (provided by the front-end via the API) and managing the appliance processes (launch, health check, restart, delete). The actions performed by the plugin run as asynchronous tasks, which CloudLaunch will instantiate and monitor (via the *deployment* endpoint). Notably, the back-end plugins were designed to be independent of the CloudLaunch framework. This is particularly interesting from the developer's perspective because it avoids developing plugins exclusively for CloudLaunch. Instead, once developed, the plugin can be used as a standalone module for handling captured application deployment actions or integrated with other deployment solutions. The opposite is also true, assuming the application deployment process has been captured using the CloudBridge library for multi-cloud compatibility, it becomes very straightforward to also integrate that application with CloudLaunch by simply implementing the defined interface.

### 3.3. Plugin Composition

Because most of the appliances will share some aspects of the interface and the launch process, we have designed the plugins to be composable. This is true of both the front-end and back-end. Users can reuse existing plugins to create more powerful components without duplicating effort. For example, the Galaxy CloudMan appliance, in addition to some optional elements, requires the user to choose the type of storage they want to use and a desired storage size. The Genomics Virtual Lab (GVL) [8] appliance represents a superset of this functionality allowing users to also launch additional services on the deployed system, for example, use of the remote desktop or Jupyter Notebook [16]. Hence, the GVL plugin is composed of the CloudMan plugin component and extended to offer the additional choices. As a result, the implementation of the GVL plugin as a complex appliance is quite straightforward counting of the order of 100 lines of code across the back-end and front-end pieces. The CloudMan plugin that implements much of the details counts approximately 300 lines of code, which itself depends on the base-plugin plugin that counts about 350 lines of code.

In addition, the front-end contains a number of reusable components that span all appliances, such as for gathering cloud credentials for users, and selecting target cloud settings. Since these components are reusable and configurable, it can significantly save on the time required to implement a new appliance plugin for CloudLaunch.

Building on this feature, we hope that this composable plugin architecture will foster an ecosystem of plugins and extensions where a community of application developers and deployers can quickly and easily assemble a more complex application out of existing plugins and integrate their application into CloudLaunch with minimal effort. As part of future work, we plan on entirely separating plugins from the CloudLaunch codebase and have them instead be self-installable plugins, which will further promote reuse of plugins and decrease dependencies on CloudLaunch.

### 3.4. Scalability

Some of the launch tasks require significant time to complete all the necessary actions, implying that the launch server resources are occupied for that duration. We have hence designed CloudLaunch to be highly scalable, with no centralized state other than a relational database for simple metadata, and is built out of well-known components with proven scalability, enabling high horizontal scaling. For example, the backend Django web-server does not use session state, and can be easily distributed over multiple servers if required. All long-running tasks are farmed out to a Celery distributed task queue, which in turn can be horizontally scaled as required. Postgres is used as the preferred database, and is a database with proven scalability characteristics. The front-end Angular application is a Single Page Application (SPA) that can be compiled entirely into static content and served in a distributed fashion through highly scalable and proven web-servers like NGINX. As a result, we do not see any obvious bottlenecks to the performance and scalability characteristics of CloudLaunch.

**Fig. 3.** User interface for launching the Base VM plugin – in this case, a Ubuntu virtual machine. Because this type of appliance requires no special input from the user, the shown elements represent the default user interface features supplied by the CloudLaunch framework for all appliances.

## 4. Demonstrations

To showcase the described features of CloudLaunch, we have implemented a number of appliance plugins. These include an appliance for launching a simple VM with a base operating system, launching an arbitrary container from Docker Hub, a complex virtual lab deployment, and several more as combinations thereof. These can be explored and launched from a live instance of CloudLaunch available at https://launch.usegalaxy.org/. Instead of using a public server, CloudLaunch can be deployed locally - the front-end (https://github.com/galaxyproject/cloudlaunch-ui) and the back-end (https://github.com/galaxyproject/cloudlaunch/tree/dev) code repositories on GitHub have required instructions. There is also an Ansible playbook available for automated deployment of the server in a production environment (https://github.com/galaxyproject/ansible-cloudlaunch). Complete implementation details for each appliance and their integration with CloudLaunch is available on GitHub: https://github.com/galaxyproject/cloudlaunch-ui/tree/master/src/app/catalog/plugins for the front-end and https://github.com/galaxyproject/cloudlaunch/tree/dev/django-cloudlaunch/baselaunch/backend_plugins for the back-end.

### 4.1. Base VM appliance

A majority of appliances deployed in a cloud environment will require that a virtual machine be provisioned. The Base VM appliance consists of a simple front-end component that implements the configName interface method, returning the desired top-level name of the appliance in the JSON response (e.g., config_ubuntu). Because no custom user interface elements are defined by the appliance, no appliance-specific data is sent to the backend. Fig. 5 shows the appliance interface (in this case just the default CloudLaunch framework interface for launching cloud appliances) while Fig. 6 captures an example data sent to the backend.

The back-end plugin receives the launch data and implements all of the details required to launch a virtual machine. These tasks include creating or reusing a key pair, creating a security group and ensuring appropriate rules are enabled, configuring private network setup, launching an instance, waiting for it to start, and associating a public IP address with the instance. As indicated earlier, all of the plugin steps run as separate asynchronous tasks, allowing the process to be of arbitrary complexity and duration.

```
{'application': 'ubuntu',
 'application_version': '16.04',
 'name': 'Demo Ubuntu launch',
 'target_cloud': 'amazon-us-east-n-virginia',
 'config_app': {
  'config_cloudlaunch': {
   'customImageID': None,
   'instanceType': 'c5.large',
   'keyPair': '',
   'network': None,
   'placementZone': '',
   'provider_settings': {
    'ebsOptimised': '',
    'volumeIOPS': ''},
   'rootStorageType': 'instance',
   'staticIP': '',
   'subnet': ''}},
 'credentials': {
  'access_key': 'AKIAJ5ZFYVIOKHZJOZBQ',
  'name': 'Galaxy Outreach Project',
  'url': 'http://<IP>/api/v1/auth/user/credentials/aws/5/'},
 'cloud': {
  'cloud_type': 'aws',
  'name': 'Amazon US East - N. Virginia',
  'region_name': 'us-east-1'}}
```

**Fig. 6.** A sample of data sent from the front-end to launch a base VM instance. The request includes all the information required to launch the selected appliance, such as the appliance version details, appliance configurations, user's credentials, and target cloud properties. In this case, no appliance-specific data is necessary so only CloudLaunch-framework data is transferred (visible under config_app → config_cloudlaunch).

**Fig. 7.** The user interface when launching the GVL platform, which offers appliance-specific launch parameters in addition to the ones for a basic VM.

## 4.2. Genomics Virtual Lab (GVL) appliance

The GVL appliance allows instances of the GVL platform [8] to be configured and launched. The GVL is a middleware layer of machine images, cloud management tools, and online services that enable researchers to build arbitrarily sized compute clusters on demand, pre-populated with fully configured bioinformatics tools, reference datasets, workflow, and visualization options; it is a complex appliance requiring a number of user options to be selected, or defaults provided. On the first page of the launch wizard (Fig. 7), the appliance user interface is the same as the base VM appliance and captures the appliance version and user cloud credentials. The second page of the wizard however, includes GVL-specific elements. The most basic ones just include the password, but the optional advanced ones (not shown for brevity) include a number of configuration options tailored for the GVL platform.

```
<div [formGroup]="form">
 <!-- GVL dashboard access password -->
 <div class='form-group'>
  <label for="id_password">
    What password would you like to use for this GVL
    instance?</label>
  <input id="id_password" type="password"
    [formControl]="gvlPasswordControl">
  <span class="material-input"></span>
 </div>
 <div class="togglebutton">
  <input type="checkbox" (click)="toggleAdvanced()">
    Advanced GVL options
 </div>
 <div [hidden]="!showAdvanced">
  <!-- CloudMan settings -->
  <config-panel>
   <cloudman-config [cloud]="cloud"
     [initialConfig]="initialConfig.config_gvl"
     [password]="gvlPasswordControl.value">
   </cloudman-config>
  </config-panel>
  <!-- Additional GVL settings -->
  <config-panel>
   <!-- GVL Component Selection -->
   <div class="form-group">
    <div class="checkbox">
     <input type="checkbox" name="gvl_cmdline_utilities"
       formControlName="gvl_cmdline_utilities" />
       GVL Commandline Utilities
    </div>
    <div class="checkbox">
     <input type="checkbox" name="smrt_portal"
       formControlName="smrt_portal" />
       PacBio SMRT Analysis
    </div>
   </div>
  </config-panel></div>
</div>
```

**Fig. 8.** The user interface implementation of the GVL appliance. The top panel shows the elements visible by default; the middle panel includes the CloudMan appliance interface (hidden under the advanced options); and the bottom panel shows the optional, advanced GVL appliance options.

On the front-end, the GVL component extends the Base VM appliance and includes the CloudMan component. It then extends those components with its own settings, specifically, the SMRT portal application and the Command Line Utilities. Despite being a complex appliance, the implementation is succinct, shown in Fig. 8. For simplicity of user experience, most of the appliance options are hidden under the advanced toggle but a key realization here is that the entire implementation of the CloudMan appliance is included here with just a few lines of code (middle panel in the figure); yet, the CloudMan appliance contains additional half dozen advanced application controls that are automatically included.

The back-end appliance plugin is also a composition of other plugins (see Fig. 9). The appliance plugin simply implements the required interface, offloading much of the complexity onto the lower-level appliances. The key aspect is to realize that the modular and composable architecture permits definition of arbitrarily complex and independent appliances. Further, building appliances on each other, the complexity can be effectively evaded through reuse.

```
class GVLAppPlugin(SimpleWebAppPlugin):
 def validate_app_config(name, cloud_version_config,
           credentials, app_config):
  gvl_config = app_config.get("config_gvl")
  user_data = CloudManAppPlugin().process_app_config(
    name, cloud_version_config, credentials, gvl_config)
  install_cmdline = gvl_config.get('gvl_cmdline_utilities',
           False)
  install_smrtportal = gvl_config.get('smrt_portal', False)
  user_data['gvl_config'] = {
    'install': [install_cmdline, install_smrtportal]}
  return user_data

 def sanitise_app_config(app_config):
  # Omitted for brevity

 def launch_app(self, task, name, cloud_version_config,
         credentials, app_config, user_data):
  ud = yaml.dump(user_data)
  result = super(GVLAppPlugin, self).launch_app(
    task, name, cloud_version_config, credentials,
    app_config, ud)
  result['cloudLaunch']['applicationURL'] = \
    'http://{0}'.format(result['cloudLaunch']['publicIP'])
  return result
```

**Fig. 9.** Implementation of the GVL appliance back-end plugin.

## 4.3. Docker Launch appliance

With an increasing number of appliances being delivered as containerized infrastructure, we have developed an appliance for launching arbitrary containers from Docker Hub (a public repository of container images). Through CloudLaunch, a user can interactively search Docker Hub, modify exposed properties of available containers through the appliance launch wizard, and launch a VM instructing it to start the chosen container (see Fig. 10). The front end of the Docker Launch appliance is a relatively complex Angular component because it requires data to be fetched from a remote service, deals with the cross-origin requests, formats the responses, and allows the user to configure image properties. However, the backend plugin is still an extension of the Base VM appliance where the only extensions include ensuring the proper firewall ports are open and the container start command is composed. Complete implementation details are available in the GitHub repository.

## 5. Related Work

CloudLaunch sits on the intersection of application deployment tools and a cloud API. Application deployment actions involve provisioning, management, and configuration of resources to make them suitable for application execution. Tools such as Ansible [6], Chef [17], Puppet [18], and SaltStack (see [19] for a comprehensive review) perform these tasks well and represent a suite of configuration management tools, with each having a slightly different approach to state and resource management. Appliance plugins integrated with CloudLaunch can internally leverage any of those tools for the relevant resource state management. CloudLaunch facilitates the entire appliance lifecycle, from deployment to deletion including resource provisioning and initial configuration but not directly state management, which can be handled internally via application plugins.

**Fig 10.** Docker Launch appliance showcasing the ability to search for arbitrary container images on Docker Hub and configure exposed image properties for editing. CloudLaunch will then launch a virtual machine, instructing it to start the selected container with specified properties.

Tools such as OpenStack Heat [20], AWS CloudFormation [21], and HashiCorp Terraform [22] focus on the entire application lifecycle via configuration files. The configuration files specify required resources and their state while the underlying framework ensures the required state is reached. These tools are more closely related to CloudLaunch and where CloudLaunch differentiates is the deployment uniformity across cloud providers. These existing tools require that appropriate configuration files be developed repeatedly for each provider. Terraform can interface with multiple clouds but requires specific resource requirements to be defined for each cloud. In contrast, appliance plugins developed for CloudLaunch operate uniformly across any supported provider without conditionals on the target cloud provider. A somewhat comparable approach is followed by Cloudify [23], which implements the TOSCA standard [12] for application deployment. TOSCA is an OASIS standard for deploying application blueprints in a manner similar to OpenStack Heat, or AWS CloudFormation. Cloudify provides an implementation of TOSCA. The main difference is that CloudLaunch follows a more imperative approach, whereas the TOSCA standard defines a declarative approach. We believe that having an imperative model allows for a declarative system to be built on top, but the reverse is not true, making CloudLaunch more flexible for systems that require a fine degree of control over what IaaS components are provisioned. CloudLaunch also does not preclude a standard such as TOSCA being implemented on top.

From the cloud API perspective, some of the tools mentioned earlier (e.g., Ansible), as well as some language-specific libraries (e.g., Apache Libcloud, jClouds), offer multi-cloud resources management capabilities via an API. A unique position for CloudLaunch is the fact it offers a language-agnostic REST API that is entirely uniform regardless of which cloud provider is accessed. While interfacing with the REST API requires lower-level requests to be handled when compared to a language-specific library, the API has the benefit of being language-agnostic. It is also foreseeable that language-specific bindings could be developed for CloudLaunch, possibly in an automated fashion via an API specification language (e.g., OpenAPI). A project most closely related to CloudLaunch from this perspective was Apache Deltacloud (http://deltacloud.apache.org/), also offering a REST API for multiple cloud providers. However, the project has been retired.

## 6. Conclusions and Future Work

As an increasing number of application become cloud-enabled, it is desirable to enable application developers and deployers to make the given applications available for launching on a variety of clouds without requiring duplicate effort. Similarly, end users should be able to discover available applications and deploy those on a cloud to which they have access while using a consistent interface. With these aims in mind, we developed and presented CloudLaunch as a web application and an API platform for discovering and launching cloud-enabled applications. Internally, CloudLaunch implements a modular and composable model allowing for arbitrary applications to be added while minimizing the amount of effort required to integrate a new appliance.

CloudLaunch has been deployed for public use since Feb 2017 and is seeing on the order of 200 appliance launches per month. Looking into the future, we intend to make it easier to integrate clouds and plugins into CloudLaunch. For plugins, we will make them standalone modules that can be dynamically loaded into CloudLaunch; similarly, we will make is possible for end-users to define their own cloud provider properties so additional resources can be readily used from the hosted service.

**Acknowledgements**

This project was supported by the National Human Genome Research Institute [grant number 5U41HG006620-06], National Cancer Institute [grant number CA184826], National eCollaboration Tools and Resources [grant number VLS402], Australian National Data Service [grant number eRIC07], and the European Regional Development Fund [grant number KK.01.1.1.01.0009].

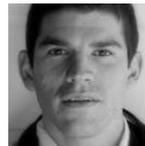
**Enis Afgan** is a research scientist in the Taylor Lab at Johns Hopkins University. He obtained his Ph.D. in 2009 from the Department of Computer and Information Sciences at the University of Alabama at Birmingham. His research interests focus around distributed computing, with emphasis on utility of Cloud Computing resources. He currently works on the Galaxy Project (galaxyproject.org) with the goal of establishing scalable, comprehensive, and accessible data analysis environments using Cloud Computing resources.

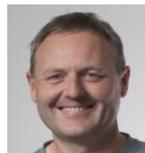
**Andrew Lonie** is director of Melbourne Bioinformatics (http://melbournebioinformatics.org.au), director of the EMBL Australia Bioinformatics Resource (EMBL-ABR: http://embl-abr.org.au), and an associate professor at the Faculty of Medicine, Dentistry and Health Sciences at the University of Melbourne, where he coordinates the MSc (Bioinformatics) and leads the Genomics Virtual Laboratory program.

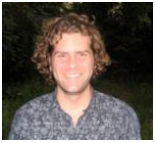 **James Taylor** is the Ralph S. O'Connor Associate Professor of Biology and Associate Professor of Computer Science at Johns Hopkins University. He is one of the original developers of the Galaxy platform for data analysis, and his group continues to work on extending the Galaxy platform. His group also works on understanding genomic and epigenomic regulation of gene transcription through integrated analysis of functional genomic data. James received a PhD in Computer Science from Penn State University.

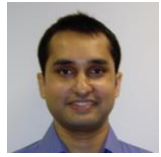 **Nuwan Goonasekera** is a research fellow at Melbourne Bioinformatics of the University of Melbourne. He obtained his Ph.D. in 2012 from the Faculty of Science and Engineering at the Queensland University of Technology in Brisbane, Australia. He currently works on the Genomics Virtual Lab (www.gvl.org.au).